\documentstyle[aps,multicol,amssymb,psfig,float,times]{revtex}

\newcommand{\be}{\begin{equation}}
\newcommand{\ee}{\end{equation}}
\newcommand{\bea}{\begin{eqnarray}}
\newcommand{\eea}{\end{eqnarray}}

\begin{document}

\draft

\title{Late-time correlators in semiclassical particle-black hole scattering}

\author{Alessandro Fabbri$^{\ast}$}
\address{Dipartimento di Fisica dell'Universit\`a di Bologna and INFN
sezione di Bologna,\\
Via Irnerio 46, 40126 Bologna, Italy.}
\author{Diego J. Navarro$^{\ast \ast}$, Jos\'e Navarro-Salas$^{\dagger}$ and Gonzalo J. Olmo $^{\ddagger}$}
\address{Departamento de F\'{\i}sica Te\'orica and IFIC, Centro Mixto
Universidad de Valencia-CSIC.\\
Facultad de F\'{\i}sica, Universidad de Valencia, Burjassot-46100, Valencia,
Spain.}

\maketitle

\begin{abstract}

We analyse the quantum corrected geometry and radiation in the
scattering of extremal black holes by low-energy neutral matter.
We point out the fact that the correlators of local observables
inside the horizon are the same as those of the vacuum. Outside
the horizon the correlators at late times are much bigger than
those of the (thermal) case obtained neglecting the backreaction.
This suggests that the corrected Hawking radiation could be
compatible with unitarity.

\end{abstract}

\pacs{PACS number(s):04.70.Dy, 04.62.+v }

\begin{multicols}{2}

\narrowtext

The discovery that black holes emit thermal radiation \cite{h1}
has raised a serious conflict between quantum mechanics and
general relativity. If a black hole is formed from the collapse of
matter, initially in a pure quantum state, the subsequent
evaporation produces radiation in a mixed quantum state \cite{h2}.
If the analysis is performed in a fixed background geometry it is
very hard to imagine how this conclusion can be avoided. The core
of the problem is connected with the black hole causal structure. The
information that flows through the horizon is not accessible to
the outside observer and therefore one has to trace over the
internal (unobserved) states. This generates a density matrix and
the information, codified in correlations between internal and
external states, is indeed lost in the singularity. There are
several posibilities to avoid such a radical conclusion, but the
most conservative one suggests that the information is recovered
in the corrected Hawking radiation due to large backreaction
effects \cite{page,tH1,tH2}. However it is difficult to unravel a
detailed mechanism capable to produce information return. Even
more, it seems unlikely that unitarity can be preserved within the
semiclassical approximation. It is usually stated that unitarity
can only be obtained in a pure quantum gravity theory.   Since we
still do not have such a theory it is useful to consider a
particular situation for which the problem can be simplified and,
in turn, the backreaction effects can be controlled in a very
efficient way. Such a scenario is given by the scattering of
low-energy particles by extremal Reissner-Nordstr\"om charged
black holes.\\
We now briefly recall the standard
picture of the process in a fixed background spacetime
approximation. Throwing long-wavelength particles into an extremal
black hole results into a non-extremal one which then emits
Hawking radiation. The Penrose diagram of such a process is given
in Fig.1. There exists radiation flowing to future null
infinity $ I^{+}$ (Hawking radiation) and in general also inside
the horizon.
The quantum state of
radiation is given by:
\be
\label{vac} |0\rangle_{in} =\sum_{i,j} c_{i,j}
|\psi_{i}\rangle_{int}\otimes|\psi_{j}\rangle_{ext} \ee i.e., a
superposition of products of internal and external states of
right-moving modes (note that we shall be mostly concerned with
right-movers,
as in \cite{GN},
because they are the ones which
transmit the Hawking radiation).
At late time this state
takes the form
\be
|0\rangle_{in}=\prod_{w}\sqrt{1-e^{-2\pi w/\kappa}}\sum_{n}e^{-\pi
nw/\kappa} |n_{w}\rangle_{int}\otimes|n_{w}\rangle_{ext} \ee where
$|n_{w}\rangle$ is a n-particle state with frequency $w$. An
observer on $I^{+}$ will describe his measurements in terms of a
reduced thermal density matrix
\be
\rho=\prod_{w}(1-e^{-2\pi w/\kappa})\sum_{n}e^{-2\pi nw/\kappa}
|n_{w}\rangle_{ext}\langle n_{w}|_{ext}. \ee

\begin{figure}
\centerline{\psfig{figure=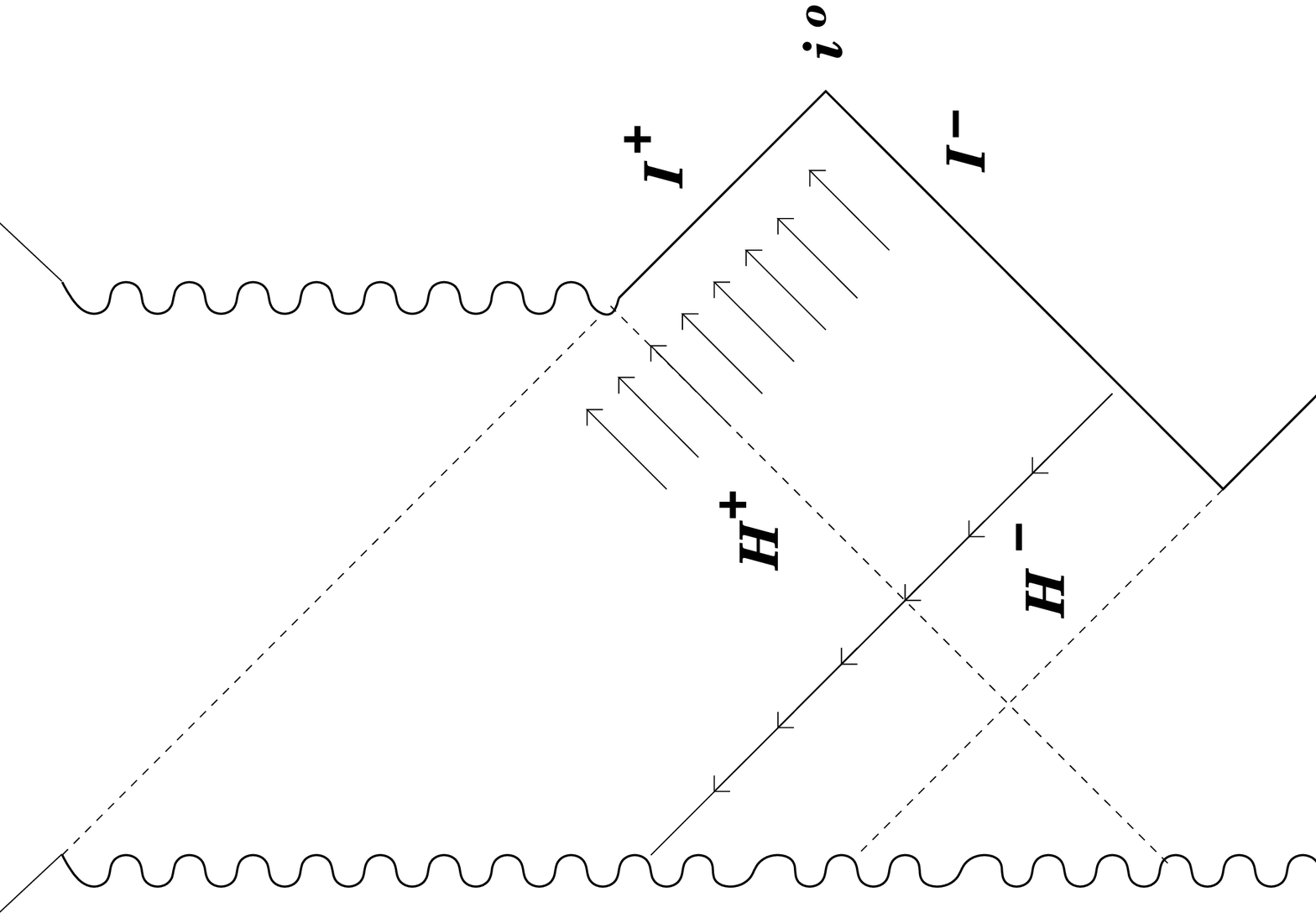,width=2.6in,height=1.7in,angle=-90}}
\end{figure}
\begin{center}
\makebox[8.5cm]{\parbox{8.5cm} {\small FIG.1.Penrose diagram
corresponding to the creation of a near-extremal charged black
hole from the extremal one. The ingoing arrow line represents an
infalling shock wave.}}
\end{center}

In this paper we shall analyse how this scenario gets modified
when backreaction effects are taken into account. Due to Hawking
emission the radiating non-extremal configuration will decay back
to the extremal black hole, if charged particles are sufficiently
massive. The corresponding Penrose diagram is given in
Fig.2. Comparing the diagrams of Figs. 1 and 2 we see that
the right singularity, being an artifact of the fixed background
approximation, has completely disappeared. It appears very
unlikely the preservation of purity if radiation is still present
at $H$ (which is part of the future Cauchy horizon), since this
would mean that the information is indeed lost in another
causally disconnected and  asymptotically flat region.\\

\begin{figure}
\centerline{\psfig{figure=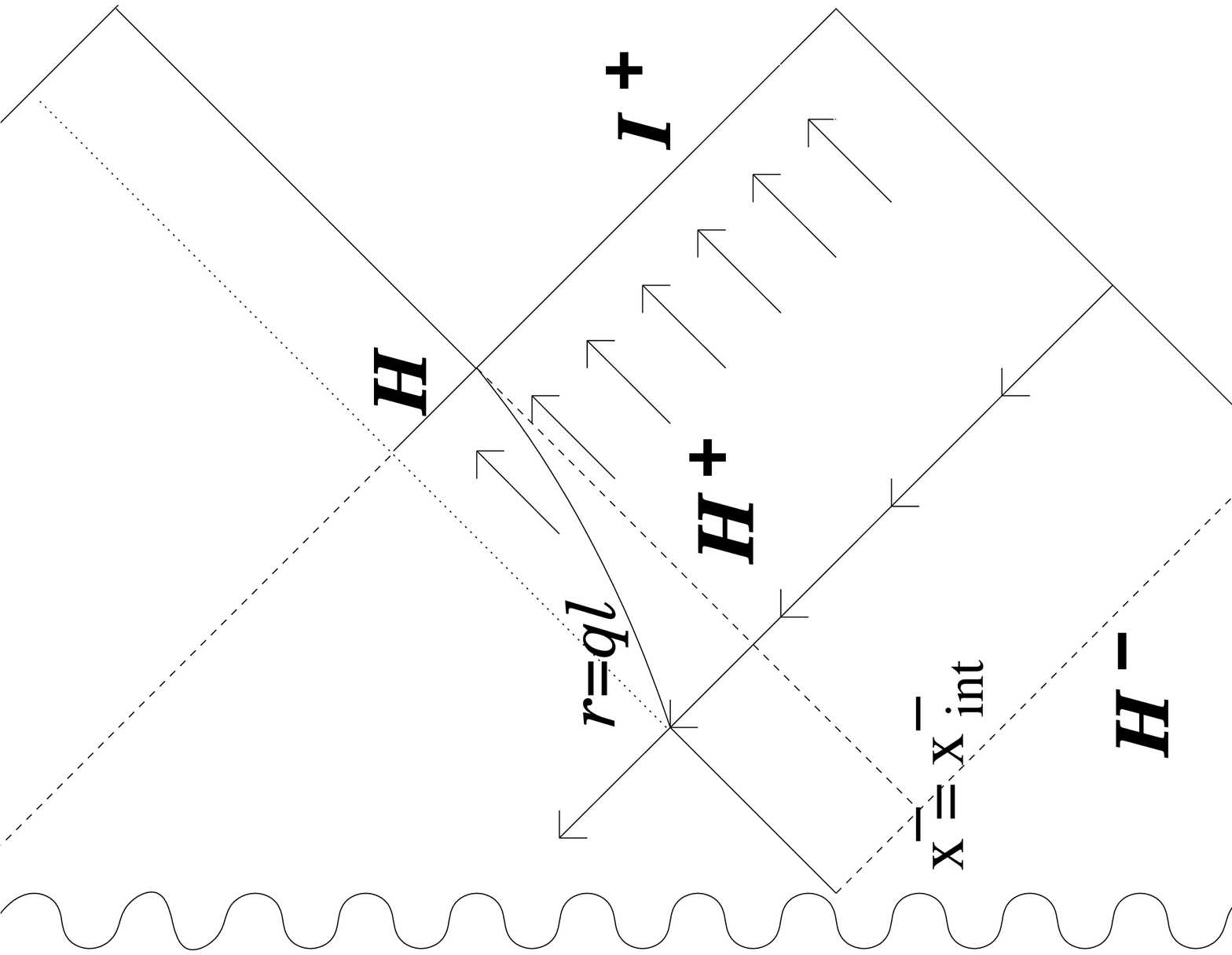,width=2.2in,height=2in,angle=-90}}
\end{figure}
\begin{center}
\makebox[8.5cm]{\parbox{8.5cm} {\small FIG.2. Penrose diagram
corresponding to the process of particle capture by an extremal
charged black hole followed by Hawking radiation. The end-state
geometry is, due to backreaction effects, an extremal black
hole. The location of the event horizon $H^{+}$ is at
$x^{-}=x^{-}_{int}$.}}
\end{center}

 We shall exploit the fact that the dominant Hawking emission is
carried away in s-waves. Moreover, in the region very close to the
initial extremal horizon $r=ql$ ($q$ is the black hole charge and
$l^{2}$ is Newton's constant), which is the relevant one to study
the radiation at $H$, a scalar matter field $f$ obeys the free
equation \be \label{field}
\partial^{2}_{t}f-\partial^{2}_{r^*}f=0,
\ee where $r^*$ is the tortoise coordinate. The dynamics in the
region close to $r=ql$ is controlled by the Jackiw-Teitelboim
model \cite{jac}, as it has been explained in \cite{f1}. The
advantage of this model is that the backreaction effects can be
incorporated immediately by adding the Polyakov-Liouville term
\cite{pol}. In summary, the effective semiclassical model is given
by the action \bea \label{paction} I &=& \int d^2x \sqrt{-g}
\left[ R \tilde{\phi} + 4 \lambda^2 \tilde{\phi} -\frac{1}{2}
|\nabla f|^2 \right] \nonumber \\ &-& \frac{\hbar}{96\pi} \int
d^2x \sqrt{-g} R \; \square^{-1} R + \frac{\hbar}{12\pi} \int d^2x
\sqrt{-g} \lambda^2 \, , \eea where the relation between the
fields appearing in (\ref{paction}) and the four-dimensional
metric is given by \bea
ds^{2}_{(4)}=\frac{ds^{2}_{(2)}}{\sqrt{\phi}}+4l^{2}\phi
d\Omega^{2} ,\ \ \phi=\frac{q^2}{4} + \tilde{\phi}, \eea and
$\lambda^{2}=l^{-2}q^{-3}$. Usually, in order to make physical sense
of the semiclassical approximation one considers a huge number, $N$, of
scalar fields. In this way the quantum gravitational corrections can be
safely neglected at one-loop order. Here for simplicity we consider $N=1$,
but it is straightforward to generalise our results to arbitrary $N$.
We note that although the model we
study is certainly simplified compared to the original 4d one the
approximations made are reasonable. Indeed, the Hawking radiation
derived from the model (\ref{paction}) has
 the same form as for 4d scalars
in the limits considered, i.e. close to the horizon and $I^+$ at late
times.
The initial extremal configuration can
be described, near $r=ql$, by the solution
 \bea
ds^{2}=-\frac{2l^{2}q^{3}dx^{+}dx^{-}}{(x^{-}-x^{+})^{2}},\ \ \
\tilde{\phi}=\frac{lq^{3}}{x^{-}-x^{+}} .\label{confmet} \eea
 The line
$x^{-}=+\infty$ corresponds to the extremal radius $r=ql$, i.e.
$\tilde{\phi}=0$. This configuration is quantum mechanically stable and it
does not
produce radiation. If we send a very narrow pulse of classical
null matter at $x^{+}=x^{+}_{0}$ with small energy $\Delta m$ we
create a near-extremal black hole of mass $m=q+\Delta m$. The
semiclassical solutions are now more involved, due to the
non-locality of the quantum effective action.

We are interested in the Hawking radiation detected by an external
observer at $I^{+}$. In this region the
quantum incoming flux vanishes and therefore the metric can be
naturally described in the outgoing Vaidya-type form

\be \label{vdy} ds^{2}=-\left(\frac{2\tilde{x}^{2}}{l^{2}q^{3}}-
l\tilde{m}(u)\right)du^{2}-2dud\tilde{x}, \ee where
$\tilde{x}=l\tilde{\phi}$ and $u$ is a null Eddington-Finkelstein
coordinate. The relevant semiclassical equations in conformal
gauge, $ds^2=-e^{2\rho}dx^{+}dx^{-} $, are

\bea \label{constraints}
-2\partial^2_{+}\tilde{\phi}+4\partial_{+}\rho\partial_{+}\tilde{\phi}
& = & -\frac{\hbar}{12\pi}\left[
(\partial_{+}\rho)^2-\partial^2_{+}\rho \right]  \\ \label{constq}
-2\partial^2_{-}\tilde{\phi}+4\partial_{-}\rho\partial_{-}\tilde{\phi}
& = & -\frac{\hbar}{12\pi}\left[
(\partial_{-}\rho)^2-\partial^2_{-}\rho \right]  \\ & &
-\frac{\hbar}{24\pi} \left(\frac{du}{dx^{-}}\right)^2\{x^{-},u\}\nonumber
\eea where $\{x^{-},u\}$ is the Schwarzian derivative proportional to
the (late time) Hawking flux. In conformal coordinates, where the
metric takes the form (\ref{confmet}), the effects of the
evaporation are all encoded in the field $\tilde{\phi}$, expressed by
means of a single function $G(x^{-})$ through

\be \label{sop}
\tilde{\phi}=\frac{G(x^{-})}{x^{+}-x^{-}}+\frac{1}{2}G'(x^{-}).
\ee The consistency of (\ref{vdy}) with the equations
(\ref{constraints}-\ref{constq}) and (\ref{sop}) implies that
$du/dx^{-}=-lq^{3}/G(x^{-})$ where $G(x^{-})$ satisfies the
differential equation

\be \label{g3} G'''= -\frac{\hbar}{24\pi}
\left(-\frac{G''}{G}+\frac{1}{2}\left(\frac{G'}{G}\right)^{2}\right).
\ee

The recovery of  the extremal solution at late times ($u\to
+\infty$) requires that $\tilde{m}(u)$ (the mass deviation from
extremality) vanishes for $x^{-}\rightarrow x^{-}_{int}$ (with
$x^{-} < x^{-}_{int}$). This implies that in this limit
$\{x^{-},u\}{\rightarrow}\ 0$, i.e.  the relation between $u$ and
$x^{-}$ is a M\"obius transformation
\be\label{eq:mob}
u=\frac{ax^{-}+b}{cx^{-}+d},
\ee
where $a,b,c,d$ are real
parameters verifying the condition $ad-bc=1$. It is now easy to
evaluate the derivative
$ du/dx^{-}=1/(cx^{-}+d)^{2}, $ and then we face two qualitatively
different possibilities: $c\neq 0$ and $c=0$. We will not consider
here the case $c=0$ as it entails a period of negative Hawking
flux (we will give more details in \cite{FNO}). Therefore the (reasonable)
assumption we make in this paper is that the Hawking radiation is always
positive.

Let us analyze the case $c\neq 0$. In Fig.3. we numerically
generate a solution for $G(x^{-})$ with this behaviour \be
\label{eq19} G(x^{-})\stackrel{x^{-}\rightarrow x^{-}_{int}}{\sim}
-\frac{1}{2}A(x^{-}-x^{-}_{int})^{2}, \ee where $A$ is a
non-vanishing constant. Note that the simplest solution which
reproduces the extremal configuration at late times is obtained
when $G(x^-)$ becomes a non-zero constant. However this implies
$c=0$. The parabolic behaviour (\ref{eq19}) is the only one which
allows to recover the extremal solution with c non-zero. Inserting
(\ref{eq19}) into eq. (\ref{sop}) we get

\be \tilde \phi = -\frac{A}{2} \frac{ (x^- - x^-_{int})(x^+ -
x^-_{int})}{x^+ - x^-} \ee which can be brought to the standard
extremal form (\ref{confmet}) after the change of coordinates
$x^{\pm '}\sim 1/(x^{\pm} - x^-_{int})$. Further, a short
manipulation of the differential equation shows that
$G^{(n)}(x^{-}_{int})=0$ for $n \geq 3$. This
implies that the unique function $G(x^{-})$, for $x^{-} \geq
x^{-}_{int}$, matching with the solution for $x^{-} \leq
x^{-}_{int}$ is exactly the parabola (\ref{eq19}). This is
crucial, since it means that inside the horizon $H^+$, and so
along $H$, we can express the solution in a form similar to
(\ref{vdy}) with $\tilde m=0$ in terms of a new null coordinate
$u_H$

\be
\label{par} u_H=-\frac{2lq^{3}}{A(x^{-}-x^{-}_{int})}.
\ee

The correlators of quasi-primary fields $\Phi_i$
associated to $f$ at $H$ are given by \cite{gin}

\bea \label{qprimf} & &\left< \Phi_1(u_{H_1}) \ldots
\Phi_n(u_{H_n})\right> = \\
& & \left(\frac{dx^{-}}{du_{H}}\right)^{\lambda_1}(x^{-}_1) \ldots
\left(\frac{dx^{-}}{du_{H}}\right)^{\lambda_n}(x^{-}_n)
 \left< \Phi_1(x^{-}_1) \ldots \Phi_n(x^{-}_n)\right> \nonumber \eea
where $\lambda_1$, \ldots ,$\lambda_n$ are the corresponding
conformal weights. Since (\ref{par}) is a M\"obius transformation
the correlators are the same as those of the vacuum. This means
that the state at $H$ is just the restriction of the vacuum to
$H$. Moreover, the range of the coordinate $u_H$ can be prolonged
beyond $H$ ($u_H \ge 0$) to cover the whole future Cauchy horizon
up to the singularity (i.e. up to $u_H\to +\infty$). This suggests
that the state inside the horizon is just the vacuum state
(naturally defined by the null time $-\infty < u_H < +\infty$)
and, therefore, that the correlators of the Hawking radiation can be 
obtained
from a pure state $|\psi\rangle_{ext}$.


\begin{figure}
\centerline{\psfig{figure=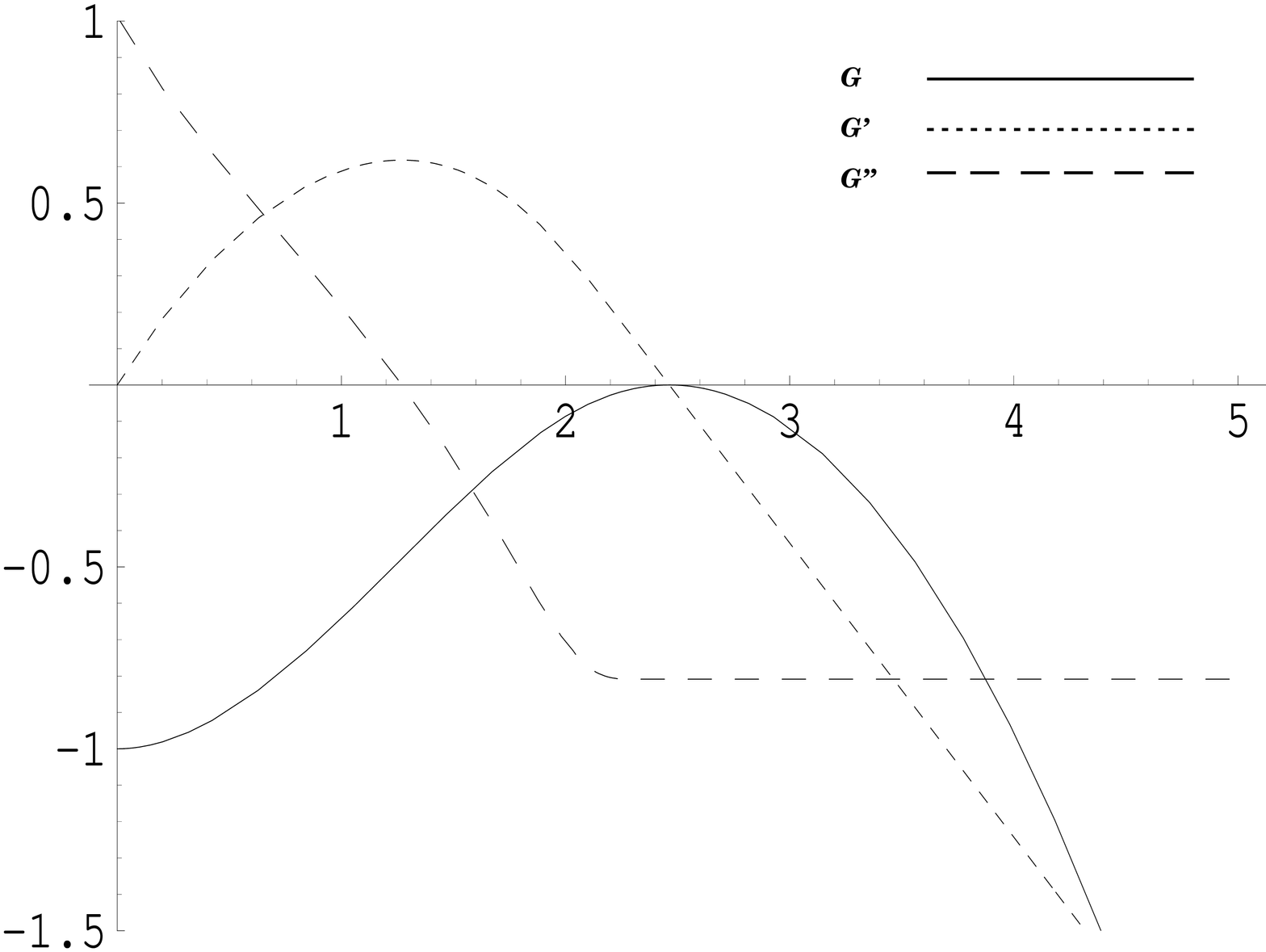,width=2.5in,height=1.7in,angle=-90}}
\end{figure}
\begin{center}
\makebox[8.5cm]{\parbox{8.5cm} {\small FIG.3.Plot of the function
G and its first and second derivatives.We have taken
$A\approx 0.808$ and $x^{-}_{int}\approx 2.463$}}
\end{center}

To deepen our analysis we shall compare the stress-tensor 2-point
correlation $C(x^{-}_1,x^{-}_2)\equiv \langle
T_{--}(x^{-}_1)T_{--}(x^{-}_2)\rangle - \langle
T_{--}(x^{-}_1)\rangle \langle T_{--}(x^{-}_2)\rangle$ measured by
the external observer at late times with and without backreaction.
It is well known that neglecting the backreaction the correlation
is thermal \be \label{tcor} C_{nb}(u_1,u_2)=
\frac{\hbar^{2}\kappa^{4}}{8\pi^{2}}\frac{e^{2\kappa|u_1-u_2|}}
{(e^{\kappa|u_1-u_2|}-1)^{4}}\ , \ee where $\kappa =\sqrt{2\Delta
m/lq^3}$ is the surface gravity at the event horizon. \\ In
general we have \cite{wil} \be
  \label{corw}
C(u_1,u_2)=\frac{\hbar^{2}}{8\pi^2}\frac{x^{-'}(u_1)^{2}x^{-'}(u_2)^{2}
}{(x^{-}(u_1)-x^{-}(u_2))^{4}}. \ee The expression (\ref{tcor}) is
obtained using the (no-backreaction) relation $x^-= -e^{-\kappa
u}/\kappa$. With backreaction effects included the relation
between $x^-$ and $u$, given by (\ref{g3}), is crucially modified
to (up to terms $O(e^{-2Cu})$)  \be
 \label{change}
 x^{-}= x^{-}_{int} -\frac{2lq^3}{Au}( 1-\frac{B}{AC} \frac{e^{-Cu}}{u} )\
,
\ee
 \\
where $C=\hbar /(24\pi lq^3)$ and $B/A= (24\pi)^2 lq^3 \Delta m/
\hbar^2 $. Therefore the two-point correlator at late times
becomes \be \label{corlate}
C_{wb}(u_1,u_2)=\frac{\hbar^2}{8\pi^2}\frac{1}{(u_1-u_2)^4}
-\frac{\hbar^2}{8\pi^2}\frac{\Delta(u_1,u_2)}{(u_1-u_2)^4}\ ,\ee
with \bea \label{dta} & &\Delta(u_1,u_2) =
  \frac{2B}{A}( e^{-Cu_1} +
e^{-Cu_2}) + \\
& &\frac{4B}{AC}( \frac{e^{-Cu_1}}{u_1} +
 \frac{e^{-Cu_2}}{u_2} +
 \frac{1}{u_1-u_2} (\frac{u_2}{u_1} e^{-Cu_1}
 -\frac{u_1}{u_2} e^{-Cu_2})) ]\ . \nonumber \eea
\\
In  the coincidence limit $u_2 -u_1 =\epsilon \rightarrow 0$
the general expression (\ref{corw})  gives
\be
C(u_1,u_2)\to \frac{1}{\epsilon^4} [ 1- \frac{8\pi}{\hbar}\left<
T_{uu}\right> \epsilon^2 ] . \ee
From eqs. (\ref{corlate}) and (\ref{dta}) it is \be C_{wb}\to
\frac{1}{\epsilon^4}[ 1-
\frac{\epsilon^2}{3}\frac{BC^2}{A}e^{-Cu_1}]\ ,\ee
from which one can extract the late time Hawking flux
$ \left< T_{uu} \right> = \frac{\hbar}{24\pi} \frac{\Delta
m}{lq^3}e^{-\frac{\hbar}{24\pi lq^3}u} $ as computed in \cite{f2}.

The increase in the correlations when backreaction effects are included
can be read off by considering the relative correlator
\be
C_{rel}\equiv \frac{C_{wb}(u_1,u_2)}{C_{nb}(u_1,u_2)}= \frac{[ 1 -
\Delta(u_1,u_2)]}{(u_1-u_2)^4}\
\frac{(e^{\kappa|u_1-u_2|}-1)^{4}}{\kappa^4 e^{2\kappa|u_1-u_2|}}\ .\ee

$C_{rel}$ by construction goes to $1$ when $u_2\to u_1$ and is
elsewhere always bigger than $1$. In particular when $\kappa|u_2 -
u_1| \gg 1$ it grows exponentially without bound. Therefore
backreaction effects restore (fully or partially) the
correlations that were lost in the (thermal) fixed background
approximation.

Summarizing, we have inspected in detail the process of particle
capture by an extremal Reissner-Nordstr\"om black hole and its
subsequent (Hawking) decay back to extremality. The solvable model
(\ref{paction}) has allowed us to determine the quantum corrected
evaporation flux as detected by an external asymptotic observer at
late times and, by analytic continuation, the quantum corrected
geometry along the future Cauchy horizon. We have given arguments
indicating that the quantum state of the radiation field in this
region is the vacuum (in particular, no radiation is present),
thus suggesting that the final state of the Hawking flux is pure
(as exemplified by the significant increase of correlations in the
emitted radiation). A full understanding of the problem requires
to construct the quantum state capable to reproduce the late time
correlator (\ref{corlate}): the first term is reproduced by the
vacuum state and the second one (with (\ref{dta})) requires a more
involved state \cite{FNO}.

To finish we would like to remark that some years ago the
particle-hole scattering was widely studied for a dilaton gravity
model \cite{cal,GN}. This raised the hope of finding a possible
resolution of the information loss paradox in a simplified
context. However additional studies showed that unitarity was not
preserved at the one-loop semiclassical level \cite{rus} (the
emergence of strong correlations has only appeared in the
subcritical regime \cite{bpp} and is crucially related to the
presence of negative energy radiation). It was
then speculated that only
higher-order corrections could restore unitarity \cite{tho,mik}.
We believe that we have provided evidence that, for
Reissner-Nordstr\"om black holes, the effects of backreaction are
stronger than for dilaton black holes, and therefore signals of
unitarity already emerge in the
semiclassical approximation.\\

This research has been partially supported by the research grants
BFM2002-04031-C02-01 and BFM2002-03681 from the Ministerio de Ciencia y
Tecnologia (Spain) and from EU FEDER funds. G.J. Olmo acknowledges
the Generalitat Valenciana
for a fellowship. A.F. thanks R. Balbinot for useful discussions.

\vspace{0.25cm}
\noindent $^{\ast}$Email address: fabbria@bo.infn.it\\
\noindent $^{\ast\ast}$Email address:jnavarro@ific.uv.es\\
\noindent $^{\dagger}$Email address: dnavarro@ific.uv.es\\
\noindent $^{\ddagger}$Email address:  gonzalo.olmo@ific.uv.es

\end{multicols}

\vspace{-2.40cm}
\begin{thebibliography}{99}

\bibitem{h1}
S.W. Hawking, {\it Comm. Math. Phys.} {\bf 43}, 199 (1975).

\bibitem{h2}
S.W. Hawking, {\it Phys. Rev.} {\bf D14}, 2460 (1976).

\bibitem{page}
D. N. Page, {\it Phys. Rev. Lett.} {\bf 44}, 301 (1980).

\bibitem{tH1}
G. t'Hooft, { \it Nucl. Phys.} {\bf B256}, 727 (1985).

\bibitem{tH2}
G. t'Hooft, { \it Nucl. Phys.} {\bf B335}, 138 (1990).

\bibitem{GN}
S. B. Giddings and W. M. Nelson,{\it Phys. Rev.} {\bf D46}, 2486(1992).

\bibitem{jac}
R. Jackiw, in {``Quantum Theory of Gravity''}, edited by S.M. Christensen
(Hilger, Bristol, 1984), p.~403; C. Teitelboim, in {\it op. cit.}, p.~327.

\bibitem{f1}
A. Fabbri, D. J. Navarro and J. Navarro-Salas, {\it Phys. Rev. Lett.} {\bf 85}, 2434 (2000); {\it Nucl. Phys.} {\bf B595} 381 (2001); K. Diba and D. A. Lowe, {\it Phys. Rev.} {\bf D65}, 024018 (2002).

\bibitem{pol}
A. M. Polyakov, {\it Phys. Lett.} {\bf B103}, 207 (1981).

\bibitem{FNO}
A. Fabbri, J. Navarro-Salas and G. Olmo, in preparation.




\bibitem{gin}
P. Ginsparg, {\it ``Applied Conformal Field Theory''}, p.1 Les Houches (1988), Ed. E. Brezin and J. Zinn-Justin.

\bibitem{wil}
F. Wilczek, hep-th/9302096.

\bibitem{f2}
A. Fabbri, D. J. Navarro and J. Navarro-Salas,{\it Nucl. Phys.} {\bf B628}, 361 (2002);
{\it Gen. Rel. Grav.} {\bf 33}, 2119 (2001).



\bibitem{cal}
C.G. Callan, S.B. Giddings, J.A. Harvey and A. Strominger, {\it Phys. Rev.} {\bf D45}, R1005 (1992).

\bibitem{rus}
J. G. Russo, L. Susskind and L. Thorlacius, {\it Phys. Rev.} {\bf D46}, 3444(1992);
{\it Phys. Rev.} {\bf D47}, 533(1993); S.W. Hawking, {\it Phys. Rev. Lett.} {\bf 69},
406 (1992).

\bibitem{bpp}
S. Bose, L. Parker and Y. Peleg, {\it Phys. Rev. Lett.}
{\bf76},861(1996).

\bibitem{tho}
L. Thorlacius, {\it Nucl.Phys.} (Proc.Suppl.) {\bf 41}, 245 (1995);S. Giddings, hep-th/9412138;A. Strominger, hep-th/9501071.

\bibitem{mik}
A. Mikovic, {\it Class. Quant. Grav.} {\bf 13}, 209 (1996); A. Mikovic and V. Radovanovic, {\it Nucl. Phys.} {\bf B481}, 719 (1996).

\end{thebibliography}
\end{document}